\documentstyle[12pt]{article}
\begin{document}
\title{Symmetric Classical Mechanics}
\author{Nuno Costa Dias\footnote{ndias@mercury.ubi.pt} \\ Jo\~{a}o Nuno Prata\footnote{jprata@mercury.ubi.pt}
 \\ {\it Grupo de Astrof\'{\i}sica e Cosmologia} \\ {\it Departamento de F\'{\i}sica - Universidade da Beira Interior}\\ {\it 6200 Covilh\~{a}, Portugal}}
\maketitle

\rightline{GATC-00-01}
\begin{abstract}
We investigate the possibility that the semiclassical limit of quantum mechanics might be correctly described by a classical dynamical 
theory, other than standard classical mechanics. Using a set of classicality criteria proposed in a related paper, we show that the time 
evolution of a set of quantum initial data satisfying these criteria is fully consistent with the predictions of a new theory of classical 
dynamics.
The dynamical structure of the new theory is given by the Moyal bracket. This is a Lie bracket that was first derived as the dynamical structure of the Moyal-Weyl-Wigner formulation of quantum mechanics. We present a new derivation of the Moyal bracket, this time in the context of the semiclassical limit of quantum mechanics and thus prove that both classical and quantum dynamics can be formulated in terms of the same canonical structure. 
\end{abstract}

\section{Introduction}

The relation between classical and quantum mechanics is beset by a number of quite subtle problems. On one side the quantum formulation of a 
given dynamical system is typically obtained by applying a quantization prescription to the classical formulation of the system. There are 
several different quantization prescriptions that one might use. However, all of them display order problems, which means that there 
is an ambiguity in the choice of the quantum system that corresponds to the original classical one.
On the other hand one expects that for an adequate choice of quantum initial data, satisfying some general conditions, the quantum formulation 
of most dynamical systems should be able to reproduce the predictions of the original classical formulation \cite{wheeler,neumann,hartle,halliwell1}. Nevertheless, it is 
still not completely clear what these general conditions should be \cite{hartle,halliwell1}. 
This has been an active field of research. Firstly because it constitutes a fundamental problem. Quantum mechanics is 
believed to provide the most fundamental description of all physical systems. Many of these systems display a classical behaviour. Therefore, 
quantum mechanics should be able to explain the emergence of a classical domain and moreover to reproduce the predictions of classical 
mechanics. Secondly, because grasping the conditions that determine the emergence of a classical domain is expected to play an 
important role in several different fields of research like for instance, quantum cosmology, quantization of closed dynamical 
systems and semiclassical gravity \cite{hartle,halliwell2,wald}, just to name a few.

This paper concerns the problem of the semiclassical limit of quantum mechanics. More precisely, we want to investigate whether the semiclassical 
limit of quantum mechanics might be correctly described by a classical dynamical structure different from ordinary classical mechanics.  
Our motivation comes from a set of results presented elsewhere \cite{nuno1}. There we were able to prove that when the quantum initial data of 
an arbitrary dynamical system satisfy a set of classicality criteria the quantum predictions will be consistent (in some precise sense) with 
the predictions of the classical formulation of the system. 
The derivation of the criteria points out an interesting fact: for a general set of quantum initial data satisfying the criteria the 
classical predictions that display the highest degree of consistency with the quantum predictions are not the ones obtained by using the standard 
formulation of classical mechanics.
In this paper the aim is to derive the dynamical framework that provides these    
new classical predictions. As a result a new theory of classical mechanics will be presented. The theory will be named symmetric classical mechanics and its properties will be studied thoroughly. In particular we will see that: i) The new theory displays a fully consistent canonical structure, ii) the quantization prescription for symmetric classical mechanics is an isomorphism between the classical and
quantum algebras of observables and is then not riddled with ordering ambiguities, iii) continuous canonical transformations and, in
particular, the time evolution are generated by unitary operators, iv) the time-evolution unitary operator is the solution of a classical
version of the Schr\"odinger equation, v) the limit of symmetric classical mechanics as $\hbar \to 0$ is standard classical mechanics and
finally, vi) we venture the possibility that symmetric classical mechanics could be trivially coupled to quantum mechanics to obtain a consistent
theory of hybrid classical-quantum dynamics.  

Most of the previous properties are a direct consequence of the fact that the dynamical structure of symmetric classical mechanics is given by the Moyal 
bracket. This is a Lie bracket that can be obtained by a deformation of the Poisson bracket \cite{vey,bayen}. It was first derived \cite{moyal} as the dynamical 
structure for the Wigner distribution function formulation of quantum mechanics \cite{wigner1,weyl1} and has been used as the starting point for a number of semiclassical 
approximation procedures \cite{smith,carruthers,lee} namely in the context of the quantum dynamics of classically chaotic systems \cite{latka,shin}. Extensive reviews of the Moyal-Weyl-Wigner formulation of quantum mechanics are given in \cite{lee,wigner2,balazs}. 
Here, instead, we derive the Moyal bracket as the dynamical structure of the semiclassical limit of quantum mechanics proving, as a by-product, that classical 
and quantum dynamics can be formulated in terms of the same bracket structure.

\section{Classicality Criteria}

In a previous paper \cite{nuno1} we developed two different classicality criteria providing a measure of the degree of classicality of an arbitrary quantum 
system. The formalism presented can be summarized in three main steps:

{\bf 1)} Let us consider an arbitrary dynamical system with $N$ degrees of freedom. Let $(q_i,p_i)$, $1=1..N$ or simply $O_i$, $i=1..2N$ be a set of 
canonical variables spanning the phase space of the system. Classical and quantum mechanics provide two alternative descriptions of the 
configuration of the system at an arbitrary time $t_0$. The classical description is given by a set of values $O_i^0=O_i(t_0)$ for the 
canonical variables plus the associated error margins $\delta_{i}$. The quantum description is given by the physical wave function 
$|\phi>$ belonging to the physical Hilbert space ${\cal H}$. The first step is to provide a measure of the consistency of these two 
descriptions. In \cite{nuno1} we proposed two such consistency criteria.
Let us review one of them:

Let $0 \le p < 1$ be an arbitrary probability. Let $M$ be some positive integer and let us consider the set of intervals of the type:
\begin{equation}
I_i (p, M) = \left[ O_i^0 - \frac{\delta_i}{(1-p)^{1/(2M)}}, O_i^0 + \frac{\delta_i}{(1-p)^{1/(2M)}} \right],
\end{equation}
associated to each classical observable $O_i$. In each of the previous intervals we can calculate the probability $p_i$ generated by the wave
function $|\phi>$, in the representation of the corresponding quantum observable $\hat O_i$:
\begin{equation}
p_i (p,M) = \sum_{a_i \in I_i (p, M) , k} |<a_i, k | \phi>|^2,
\end{equation}
where $|a_i, k>$ is the general eigenvector of $\hat O_i$ with eigenvalue $a_i$ and degeneracy index $k$. For given values of the classical and 
quantum data $O_i^0$, $\delta_i$ and $| \phi>$, this probability is an exclusive function of $p$ and $M$. We can now state the consistency 
criterion: 

\underline{{\bf Definition 1}} {\bf - Consistency Criterion}\\
The classical and quantum data, describing a given configuration of the dynamical system will be $M$-order consistent if and only if 
for all $0 \le p <1$ and all $i=1, \cdots, 2 N$ the condition $p_i (p,M) \ge p$ is satisfied,
i.e.;
\begin{equation}
\sum_{a_i \in I_i (p, M) , k } |< a_i, k | \phi >|^2 \ge p, \qquad \forall p \in \left[0, 1 \left[ \right. \right., \forall i=1, \cdots , 2N,
\end{equation}
where $I_i (p, M)$ is given by eq.(1).
\\

This criterion provides a measure of how peaked the wave function is - in the representation of each of the quantum observables - around
the classical error margin of the corresponding classical observable. Notice that, in particular, the higher the degree of consistency the 
bigger the probability that a quantum measurement provides a value inside the corresponding classical error interval.

{\bf 2)} The second and main step in developing the classicality criterion was the derivation of the following expansion:
\begin{eqnarray}
\hat{A}-A & = & 
\sum_{i=1}^{2N} \frac{\partial A}{\partial O_{i}}(\hat{O}_{i}-
O_{i})
+\frac{1}{2} \sum_{i,j=1}^{2N} \frac{\partial^2 A}{\partial O_{i}
\partial O_{j}}(\hat{O}_{i}-O_{i})(\hat{O}_{j}-O_{j})+....
\end{eqnarray}
where $\hat{A}$ is a general operator, $\hat{A}=F(\hat{O}_i)$ and $A$ is some classical version (which version is yet to be discussed) of 
$\hat{A}$, $A=G(O_i)$.
If the expansion (4) is valid then we can easily obtain the following relation:
\begin{eqnarray}
& & <E^m(\hat{A},\phi,A^0)|E^m(\hat{A},\phi,A^0)> \le \\
& & \le \sum_{i_1=1}^{2N}...
\sum_{i_m=1}^{2N}\sum_{j_1=1}^{2N}...\sum_{j_m=1}^{2N} \prod_{k=1}^m \prod_{s=1}^m
\left. \frac{\partial A}{\partial O_{i_k}} \right|_{O_{ik}=O_{ik}^0}\left.\left( \frac{\partial A}{\partial O_{j_s}}
\right)^{\ast}\right|_{O_{js}=O_{js}^0} <E_{O_{i1},...,O_{im}}|E_{O_{j1},...,O_{jm}}>
+... \nonumber
\end{eqnarray}
where $A^0=G(O_i^0)$ and $|E^m_X>=|E^m(\hat{X},\phi,X^0)>$ is named the mth-order error ket of the operator $\hat{X}$ around the classical value $X^0$ and is given by
$|E_X^m>=(\hat{X}-X^0)^m|\phi>$ for all $m \in {\cal N}$. 
Moreover, $|E_{O_{j1},...,O_{jm}}>=(\hat{O}_{j_1}-O^0_{j_1})....(\hat{O}_{j_m}-O^0_{j_m})|\phi>$ is named the $m$-order mixed error ket.
The most important property of the error ket framework is the following: if 
$<E_X^m|E_X^m> \le \Delta^{2m}$ then, in the representation of $\hat{X}$ the wave function $|\phi>$ has at least a probability $p$ confined to 
the interval $I_m=[X^0-\frac{\Delta}{(1-p)^{1/2m}}, X^0+\frac{\Delta}{(1-p)^{1/2m}}]$. That is, $|\phi>$ is m-order consistent with the 
classical interval $[X^0-\Delta,X^0+\Delta]$ (check for the consistency criterion in step 1). The reader should refer to \cite{nuno1} for a 
detailed discussion of the error ket formalism.

This property points out a quite obvious way of developing a classicality criterion.
Let us assume for the moment that (4) is valid for $\hat{A}=\hat{O}_i(t)$ and $A=O_i(t)$, $i=1..2N$ (where $O_i(t)$ is some classical version of $\hat{O}_i(t)$ - see point 3)) and let us consider all the sequences of observables $S_{i_k}=O_{i_1},....,O_{i_n}$ associated to the sequences of values $i_k=i_1,..,i_n \in \{1..2N\}$ such that:
\begin{equation}
\frac{\partial A}{\partial S_{ik}} = \frac{\partial^n A}{\partial O_{i1}...\partial O_{in}} \not= 0,
\end{equation}
for some classical observable $A=O_i(t), i=1..2N$. Let also $S_{ik}^m$ be an array of $m$ arbitrary sequences $S_{ik}$. Moreover, we define $\delta_{S_{ik}^m}$ to be the
product of all error margins associated with the observables included in $S_{ik}^m$. We proposed the set of relations,
\begin{equation}
<E_{S_{ik}^m}|E_{S_{ik}^m}> \le \delta^2_{S_{ik}^m}, \qquad \forall m \le M,
\end{equation}
as a classicality criterion. Notice that given the classical initial data $(O_i^0,\delta_i)$, the set of inequalities (7) constitute a set of conditions (classicality conditions) on the functional form of the initial data wave function $|\phi>$. If a certain dynamical system with some given initial data satisfies these relations up to order $m=M$, then we 
say that the system is $M$-{\it order classical}. In this case it is straightforward 
to obtain from (5) that:
\begin{equation}
<E^M_A|E^M_A> \le \left(\sum_{i=1}^{2N} \left| \frac{\partial A}{\partial O_i} \right|_{O_i=O_i^0}
\delta_{i}+...\right)^{2M} = \delta_{A}^{2M}
\end{equation}
and thus we can state that if $|\phi>$ satisfies (7) up to order $M$ then 
not only $|\phi>$ is $M$-order consistent with the initial time classical description, but also all future classical and quantum descriptions of the configuration of the system are $M$-order consistent.

{\bf 3)} The key step to obtain the classicality criterion was the derivation of the  relation (4). Given $\hat{A}$ it is clear that this relation 
is not valid for an arbitrary $A$. In fact $A$ should be obtained from $\hat{A}$ by following a well defined procedure that was named 
dequantization. In \cite{nuno1}, we proved that if this dequantization is the inverse of the Dirac quantization map \cite{dirac1,dirac2}, then the general
relation (4) is not exactly valid, being however a {\it very} good approximation. More precisely, we proved that for $A$ obtained from the
operator $\hat A$ by applying the Dirac dequantization map, the difference between the right- and left-hand sides of eq.(4) is, at most
proportional to a factor of $\hbar^2$. 

On the other hand, if $\hat A$ in eq.(4) is the time-evolution of some initial observable $\hat O_i (0)$ 
(i.e. $ \hat A = \hat O_i (t)$ ), then $A$, the observable obtained by applying the Dirac dequantization map to $\hat A$, is just the time evolution of the classical
observable $ O_i (0)$, that is $A = O_i (t)$. To proceed, let us define the semiclassical limit of the quantum operator $\hat A$ to be the
classical observable $ A_S$ that fully validates the expansion (4).
 We can now restate our statement made in the previous paragraph in the following terms: The observable $O_i
(t)$ is not the semiclassical limit of the quantum observable $\hat O_i (t)$ (i.e. $O_i(t)\not= A_S$) and hence does not allow for the exact statement that if the
classical and quantum initial data satisfy the set of relations (7), then the classical and quantum predictions will be $M$-order consistent
at all times (8). Still, $O_i (t)$ provides a prediction that diverges from the correct semiclassical limit one by a factor proportional to $\hbar^2$ at
most (i.e. $O_i(t)=A_S+{\cal O}(\hbar^2)$). This means that the standard classical prediction is well inside the error interval associated with the semiclassical limit prediction.
  
Consequently, the results of \cite{nuno1} allow us to conclude that: i) Classical mechanics does not provide the exact semiclassical limit for 
quantum systems with a set of initial data satisfying (7). ii) Still, classical mechanics provides a {\it very} similar prediction to the 
exact semiclassical limit one.

In this paper we want to obtain the exact form of the classical observables that fully validates the expansion (4). More important, the aim is to 
derive a consistent dynamical framework able to provide those semiclassical limit observables directly from the classical initial data.

\section{Symmetric Dequantization}

The complete set of fundamental hermitian operators $S=\left\{\hat O_i, i=1, \cdots, 2N \right\}$ spans the algebra $\hat{{\cal A}}$ of operators acting on the Hilbert
space ${\cal H}$,
\begin{equation}
\hat{{\cal A}} \equiv \left\{ \hat A: \hat A = \sum_{i=1}^n c_i \prod_{j=1}^m \hat O_{ij}; \qquad n,m \in {\cal N}, c_i \in {\cal C} \right\},
\end{equation}
where $\hat O_{ij} \in S$. Consequently, the set
\begin{equation}
\hat{{\cal B}} \equiv \left\{\hat O_{i_1 i_2 \cdots i_k} \equiv \hat O_{i_1} \hat O_{i_2} \cdots \hat O_{i_k}; \qquad 1 \le i_1,i_2, \cdots , i_k \le 2 N, k\in {\cal N}
\right\}
\end{equation}
generates all the elements of $\hat{{\cal A}}$. We stress the fact that the order of the operators in the product $\hat O_{i_1 i_2 \cdots i_k}$ is meaningful.

$\hat{{\cal A}}$ is thus an infinite dimensional complex vector space together with the bracket rule,
\begin{equation}
\left[ \hat A, \hat B \right] \equiv \hat A \hat B - \hat B \hat A, \qquad \hat A, \hat B \in \hat{{\cal A}}.
\end{equation}
The set of fundamental operators $S$ is decomposed into $N$ pairs of variables, $(\hat q_i, \hat p_i)$, with $i=1, \cdots , N$ and:
\begin{equation}
\begin{array}{l l}
\hat O_i = \hat q_i, & i=1, \cdots, N,\\
\hat O_i = \hat p_{i-N}, & i=N+1, \cdots , 2N.
\end{array}
\end{equation}
These pairs are canonically conjugate in the sense that:
\begin{equation}
\begin{array}{l l r}
\left[\hat q_i, \hat q_j \right] = & \left[\hat p_i, \hat p_j \right] = 0; & \left[\hat q_i, \hat p_j \right] = i \hbar \delta_{ij}.
\end{array}
\end{equation}
Let $T^* M$ be the phase space of the corresponding classical system and ${\cal A}$ the algebra of classical observables, ${\cal A} \equiv 
\left\{ f \in C^{\infty}: T^* M \to  {\cal C} \right\}$. Let us define the dequantization map, $ V$: $ \hat{{\cal A}} \to  {\cal A}$, such 
that $V \circ \Lambda =1$, where $\Lambda$ is the Dirac quantization map \cite{dirac1,dirac2}. This map attributes a classical variable to every quantum operator. 

Let then $A = V ( \hat A)$. We saw in \cite{nuno1} that if $A$ is obtained from $\hat{A}$ using the map $V$ then, in general, the expansion (4) is not exactly valid, the
right and left hand sides displaying a difference of the order of $\hbar^2$ at most.  Consequently, all subsequent results, including eq.(8) display an imprecision of the
order of $\hbar^2$. Therefore the conclusion that if a given set of classical and quantum initial data are $M$-order classical, then their time
evolution will always be $M$-order consistent, is only valid up to a correction - of the classical prediction - by a term proportional to $\hbar^2$. Moreover, the map $V$ is beset by order problems: it is neither univocous nor injective, as
illustrated by the two following examples: let $\hat{q}$, $\hat{p}$ be the two fundamental operators of a one dimensional quantum system and consider the two following hermitian
operators:  $\hat{A}=\hat{q}\hat{p}^2\hat{q}$ and  $\hat{B}=1/2(\hat{q}^2\hat{p}^2+\hat{p}^2\hat{q}^2)$. Clearly, $\hat{A}\not=\hat{B}$ and yet one possible Dirac dequantization yields: 
$$
V(\hat{A})=q^2p^2 \quad \mbox{and} \quad V(\hat{B})=q^2p^2
$$
and thus the map $V$ is not injective. Using the same example we also have: $\hat{A}=\hat{B}+\hbar^2$ and yet:
$$
V(\hat{A})=q^2p^2 \not= V(\hat{B}+\hbar^2)=q^2p^2+\hbar^2
$$
and so the map $V$ is not univocous either.

We see that there are many different ways of dequantizing a system. This should come as no surprise since the Dirac quantization map is itself
not one to one. As a consequence the classical algebra ${\cal A}$ which is spanned by:
\begin{equation}
{\cal B} \equiv \left\{ O_{i_1 i_2 \cdots i_k} \equiv  O_{i_1}  O_{i_2} \cdots  O_{i_k}; \qquad 1 \le i_1,i_2, \cdots , i_k \le 2 N, 
k\in {\cal N} \right\},
\end{equation}
where this time the order of the observables is immaterial, shows no 
straightforward relation with $\hat{{\cal B}}$.

The whole framework described in the last section (and in particular the expansion (4)) is exactly valid if the classical observables $A$ are
obtained (through a trivial substitution of the fundamental observables $\hat O_i$ by the classical ones $O_i$) from a fully symmetric form of
$\hat A$ \cite{nuno1}. Consequently, the aim now is to present a dequantization map that, given an arbitrary quantum observable $\hat A$, yields
the classical observable $A_S$. 

Let us thus introduce the "{\it symmetric dequantization}" prescription. Any operator in the algebra $\hat{{\cal A}}$ can be cast in the form of 
a linear combination of fully symmetrized polynomia of the fundamental operators. Consider e.g. $\hat p \hat q$:
\begin{equation}
\hat p \hat q = \frac{1}{2} (\hat p \hat q + \hat q \hat p) + \frac{1}{2} \left[ \hat p, \hat q \right]  \equiv (\hat p \hat q)_+ + \frac{1}{2} \left[\hat p, \hat q \right] ,
\end{equation}
where $(\hat p \hat q)_+ \equiv 1/2 ( \hat p \hat q + \hat q \hat p)$ is the completely symmetrized product. Since $\left[ \hat p, \hat q \right] = -i \hbar $ is a c-number, we have
achieved the expansion of $\hat p \hat q$ in the basis of the fully symmetrized operators.

H.Weyl was the first to suggest that any operator could be expanded as a sum of completely symmetric terms, \cite{weyl2}. He proposed the following rule. A given operator
$\hat b (\hat{\vec q}, \hat{\vec p})$ is represented in
the form:
$$
\hat b (\hat{\vec q}, \hat{\vec p}) = \int d \vec x d \vec y \beta (\vec x, \vec y) \exp \left( i \vec x \cdot \hat{ \vec q} + i \vec y \cdot \hat{\vec p} \right),
$$
where $\vec x$ and $\vec y$ are two $N$-dimensional vectors whose components are c-numbers, $\hat{ \vec q} = (\hat q_1, \cdots , \hat q_N)$, 
$\hat{ \vec p} = (\hat p_1, \cdots , \hat p_N)$ and $\beta (\vec x, \vec y) = \beta^* (- \vec x, - \vec y)$ is some numerical function
(possibly singular). The previous condition ensures the hermiticity of the operator. If the position $\hat q_j$ appears $n$ times and the momentum $\hat p_k$ appears $m$ times in a given operator, then we 
include terms of the form\footnote{$\delta^{(n)} (x)$ is the $n$-th derivative of the delta-function with respect to its argument.} $i^n 
\delta^{(n)} (x_j)$ and $i^m \delta^{(m)} (y_k)$, respectively, in the function $\beta (\vec x, \vec y)$. We equally include factors
$\delta (x_l)$ and $\delta (y_s)$ for all $\hat q_l$ and $\hat p_s$ that are absent in the operator. 

As an example, consider a system with one degree of freedom ($N=1$) and some operator where $\hat q$ appears twice and $\hat p$ once. From the
previous rule we get the completely symmetric operator:
$$
\hat b (\hat q, \hat p) =\int dx dy \left[ i^2 \delta ^{''} ( x) \right] \left[ i \delta^{'} (y) \right] \exp \left( i x \hat q + i y \hat p \right) = \frac{1}{3} 
\left( \hat q^2 \hat p + \hat q \hat p \hat q + \hat p \hat q^2 \right).
$$

There is a clear advantage in using the symmetric version of quantum mechanics. The set $\hat{{\cal B}}_+$ of symmetrized products,
\begin{equation} 
\hat{{\cal B}}_+ \equiv \left\{ \left(\hat O_{i_1 i_2 \cdots i_k} \right)_+ \equiv  \left( \hat O_{i_1}  \hat O_{i_2} \cdots 
\hat O_{i_k} \right)_+ ; \qquad 1 \le i_1,i_2, \cdots , i_k \le 2 N, k\in {\cal N} \right\},
\end{equation}
constitutes a basis for $\hat{{\cal A}}$. In fact it is easy to check that its elements are linearly independent and moreover that all elements
of $\hat{{\cal A}}$ can be expanded in terms of the elements of $\hat{{\cal B}}_+$. On the other hand, in the completely symmetric product $\left(
\hat O_{i_1} \hat O_{i_2} \cdots \hat O_{i_k} \right)_+$ the order of the operators is immaterial and therefore $\hat{{\cal B}}_+$ and 
${\cal B}$ have the same number of elements. As a side remark we conclude that $\hat{{\cal B}}$ is overcomplete. We can thus define the
dequantization map $V_S$ in the following way:

\underline{{\bf Definition 2}} - {\bf Symmetric Dequantization}\\
The dequantization map $V_S:$ $\hat{{\cal A}} \to {\cal A}$ is defined by the following rules: 
\\
1) $V_S$ is a linear map;\\
2) $V_S$ maps the identity to the identity;\\
3)  $V_S \left( \left( \hat O_{i_1} \hat O_{i_2} \cdots \hat O_{i_k} \right)_+ \right) =  O_{i_1}  O_{i_2} \cdots O_{i_k}, \qquad \mbox{for all }  \left(
\hat O_{i_1} \hat O_{i_2} \cdots \hat O_{i_k} \right)_+ \in \hat{{\cal B}}_+  $. 

\bigskip

Let us study some of the properties of $V_S$:\\
a) For a generic operator $\hat A$ we have:
\begin{equation}
V_S( \hat A) \equiv A = V_S \left( \sum_{i=1}^n c_i \left( \prod_{j=1}^m \hat O_{ij} \right)_+ \right) = \sum_{i=1}^n c_i  \prod_{j=1}^m  O_{ij},
\end{equation}
where $\sum_{i=1}^n c_i \left( \prod_{j=1}^m \hat O_{ij} \right)_+$ is the expansion of the operator $\hat A$ in the basis $\hat{{\cal B}}_+$.\\
b) $V_S \left( \left[ \hat O_i, \hat O_j \right] \right) = i \hbar \left\{ O_i, O_j \right\}$, where  $\left\{ \quad, \quad \right\}$ is the Poisson bracket defined by:
\begin{equation}
\left\{A, B \right\} \equiv \sum_{i=1}^N \left[ \frac{\partial A}{\partial q_i} \frac{\partial B}{\partial p_i} -\frac{\partial A}{\partial p_i} \frac{\partial B}{\partial 
q_i} \right], \qquad  \forall A,B \in {\cal A}
\end{equation} \\
c) From the expansion (17) and the definition of $V_S$, we conclude that:
\begin{equation}
V_S(\hat A^{\dagger}) =\left[V_S(\hat A) \right]^* \equiv A^*.
\end{equation}\\
d) Since the two basis $\hat{{\cal B}}_+$ and ${\cal B}$ have the same number of elements, the map $V_S$ is bijective and univocous.\\
e) From property a), we see that $A= V_S(\hat A)$ is the appropriate classical observable to be used in the expansion (4) of section 2. In that case the expansion (4) is
exactly valid. Consequently, all the results of section 2 concerning the consistency between the classical and the quantum description of the time evolution of a general
observable $\hat A$ become exactly valid. \\
f) In general $V_S( \hat A \hat B) \ne V_S(\hat A) V_S (\hat B)$ and $V_S \left( \left[ \hat A, \hat B \right] \right) \ne i \hbar \left\{V_S(\hat A), V_S(\hat B) \right\}$.
Let us then define a new product and a new bracket in ${\cal A}$:

\underline{{\bf Definition 3}} - {\bf The product $*$ and the bracket $\left[ , \right]_M$}\\
The new product is the map:
\begin{equation}
* : {\cal A} \times {\cal A} \to {\cal A}; \qquad V_S( \hat A) * V_S( \hat B) = V_S( \hat A \hat B), \qquad \forall \hat A, \hat B 
\in \hat{{\cal A}}.
\end{equation}\\
Using this product we can define the new classical bracket:
\begin{equation}
\left[  , \right]_M: {\cal A} \times {\cal A} \to {\cal A}; \qquad \left[ A , B  \right]_M = A * B - B * A,
\end{equation}
and it is straightforward to check that the new bracket satisfies the identity: $\left[V_S( A) , V_S(B)\right]_M = V_S \left( \left[ \hat A , \hat B \right] 
\right)$.
 
The aim of the next section is to derive the explicit formula of the product $V_S( \hat A \hat B)$ and of the commutator $V_S( \left[ \hat A, \hat B \right])$ and, in the
sequel, study their properties. We will see that the product $*$ is well defined: $({\cal A},+,*)$ is a ring and the bracket $\left[, \right]_M$ is a true 
Lie bracket. We can thus
anticipate the last property of the symmetric dequantization map:\\
g) The dequantization map $V_S$ is an isomorphism between the Lie algebras $(\hat{{\cal A}}, \cdot, \left[ , \right])$ and $({\cal A}, *, \left[, \right]_M)$.

\section{Dynamical structure of Symmetric Classical Mechanics}

The purpose of this section is thus to derive an explicit formula for the dequantization of the product $V_S( \hat A \hat B)$ of two operators 
$\hat A, \hat B \in \hat{{\cal A}}$ and, as a by-product, of the commutator $V_S( \left[ \hat A, \hat B \right])$. This will allow us to
establish the canonical structure of symmetric classical mechanics and therefore to make predictions about the evolution of an arbitrary
classical system without having to refer to its quantum formulation.

\subsection{The product $*$}

First of all, notice that performing a symmetrization in the sense of section 3, coincides with the procedure of normal ordering bosonic fields by applying Wick's
theorem, providing we define the "{\it propagator}":
\begin{equation}
< \hat p_i \hat q_j > = - < \hat q_j \hat p_i > \equiv \frac{1}{2} \left[ \hat p_i, \hat q_j \right] = -i \frac{\hbar}{2} \delta_{ij}.
\end{equation}
The following example illustrates this method:
$$
\begin{array}{c}
\hat O_i \left( \hat O_j \hat O_k \right)_+ = \left(\hat O_i \hat O_j  \hat O_k \right)_+ + <\hat O_i \hat O_j > \hat O_k + <\hat O_i \hat O_k > \hat O_j =\\
\\
= \left(\hat O_i \hat O_j  \hat O_k \right)_+ + \frac{1}{2} \left[\hat O_i , \hat O_j \right] \hat O_k + \frac{1}{2} \left[\hat O_i, \hat O_k \right] \hat O_j ,
\end{array}
$$
where $\left(\hat O_i \hat O_j  \hat O_k \right)_+ \equiv \frac{1}{3 !} \left(\hat O_i \hat O_j  \hat O_k + \mbox{permutations} \right) $. We stress that this identity only
holds, because $\left[\hat O_i , \hat O_j \right]$ is a c-number.

Let us now try to obtain $V_S( \hat A \hat B)$ and $V_S( \left[ \hat A, \hat B \right])$ for generic operators $\hat A, \hat B \in \hat{{\cal A}}$, given $A=V_S( \hat A)$,
$B=V_S( \hat B)$. If we assume that $\hat A$ and $\hat B$ are already completely symmetrized, then we get (cf.(4)):
\begin{equation}
\left\{
\begin{array}{l}
\hat A = \sum_{n=0}^{\infty} \frac{1}{n !} \sum_{1 \le i_1, \cdots , i_n \le 2N} \frac{\partial^n A}{\partial O_{i_1} \cdots \partial O_{i_n}} \left( \hat M_{i_1} \cdots \hat
M_{i_n} \right)_+\\
\\
\hat B = \sum_{m=0}^{\infty} \frac{1}{m !} \sum_{1 \le j_1, \cdots , j_m \le 2N} \frac{\partial^m B}{\partial O_{j_1} \cdots \partial O_{j_m}} \left(\hat M_{j_1} \cdots 
\hat M_{j_m} \right)_+,
\end{array}
\right.
\end{equation}
where the monomials $\hat M_i$ are defined by: $\hat M_i \equiv \hat O_i - O_i$.

The subtlety resides in noticing that in general $(\hat A \hat B)_+ \ne 1/2 (\hat A \hat B + \hat B \hat A)$. So the whole problem reduces to symmetrizing $\hat A \hat B$
properly.
If we carry out the product of the two expansions (23), we shall have to dequantize terms of the form:
\begin{equation}
V_S \left( (\hat M_{i_1} \cdots \hat M_{i_n})_+ (\hat M_{j_1} \cdots \hat M_{j_m})_+ \right).
\end{equation}  
Using Wick's theorem we have:
\begin{equation}
\begin{array}{c}
(\hat M_{i_1} \cdots \hat M_{i_n})_+ (\hat M_{j_1} \cdots \hat M_{j_m})_+ = \left(\hat M_{i_1} \cdots \hat M_{i_n} \hat M_{j_1} \cdots \hat M_{j_m} \right)_+ + \mbox{terms
with one contraction}+\\
\\
+ \mbox{terms with two contractions} + \cdots + \mbox{terms with $min \left\{n,m \right\}$ contractions}.
\end{array}
\end{equation} 
On the other hand,
$$
\begin{array}{c}
V_S \left( (\hat M_{i_1} \cdots \hat M_{i_k})_+ \right) = V_S (\hat M_{i_1}) \cdots V_S (\hat M_{i_k})=  V_S (\hat O_{i_1} -O_{i_1}) \cdots V_S (\hat O_{i_k} - O_{i_k})=\\
\\
=  ( O_{i_1} -O_{i_1}) \cdots ( O_{i_k} - O_{i_k})= 0
\end{array}
$$
Consequently, if $n \ne m$ then the term (24), (25) will yield a vanishing contribution to $V_S( \hat A \hat B)$. We are left with:
$$
\begin{array}{c}
V_S( \hat A \hat B) = \sum_{n=0}^{\infty} \frac{1}{(n !)^2} \sum_{1 \le i_1, \cdots , i_n \le 2N} \sum_{1 \le j_1, \cdots , j_n \le 2N} \frac{\partial^n A}{\partial O_{i_1}
 \cdots \partial O_{i_n}} \frac{\partial^n B}{\partial O_{j_1} \cdots \partial O_{j_n}} \times \\
\\
\times V_S \left[ <\hat M_{i_1} \hat M_{j_1}><\hat M_{i_2} \hat M_{j_2}>  \cdots <\hat M_{i_n} \hat M_{j_n}> +  \right. \\
\\
\left. +  <\hat M_{i_1} \hat M_{j_2}><\hat M_{i_2} \hat M_{j_1}>  \cdots 
<\hat M_{i_n} \hat M_{j_n}> + \mbox{permutations} \right],
\end{array}
$$ 
where the expression inside the bracket includes all the terms with $n$ contractions. Since the derivative $(\partial^n B)/(\partial O_{j_1} \cdots \partial O_{j_n})$ is symmetric with respect
to swapping any two indices, we conclude that all the terms yield the same contribution. There are $n !$ permutations of the $j$ indices. We hence get:
\begin{equation}
\begin{array}{c}
V_S( \hat A \hat B) = \sum_{n=0}^{\infty} \frac{1}{n !} \sum_{1 \le i_1, \cdots , i_n \le 2N} \sum_{1 \le j_1, \cdots , j_n \le 2N} \frac{\partial^n A}{\partial O_{i_1}
 \cdots \partial O_{i_n}} \frac{\partial^n B}{\partial O_{j_1} \cdots \partial O_{j_n}}  \times \\
\\
 \times V_S \left( \frac{1}{2^n} \left[\hat M_{i_1} , \hat M_{j_1} \right] \left[\hat M_{i_2} , \hat M_{j_2} \right]   \cdots \left[ \hat M_{i_n} , \hat M_{j_n} \right] \right).
\end{array}
\end{equation}
Notice that $V_S \left( \left[ \hat M_i, \hat M_j \right] \right) = V_S \left( \left[ \hat O_i - O_i , \hat O_j - O_j \right] \right) = i \hbar \left\{ O_i , O_j \right\}$.
Let us now define the derivative ${\buildrel { \leftrightarrow}\over\partial}$ which obeys the antisymmetric Leibnitz rule:
\begin{equation}
\begin{array}{l l}
{\buildrel { \leftrightarrow}\over\partial\! (AB)}  = & (\partial A) B - A \partial B,\\
{\buildrel { \leftrightarrow}\over\partial\! \left\{A,B \right\}} = & \left\{\partial A, B \right\} - \left\{ A, \partial B \right\}.
\end{array}
\end{equation}
We equally define the following "{\it Liouvillian}" operator;
\begin{equation}
\hat{{\cal L}} \equiv \frac{1}{2} \sum_{i=1}^N \left( \frac{\partial}{\partial p_i} \frac{ {\buildrel { \leftrightarrow}\over\partial}}{\partial q_i} 
-  \frac{\partial}{\partial q_i}  \frac{{\buildrel { \leftrightarrow}\over\partial}}{\partial p_i} \right).
\end{equation}
We represent eq.(26) in the form,
\begin{equation}
V_S( \hat A \hat B) = \sum_{n=0}^{\infty} {\cal L}_n.
\end{equation}
Let us now prove by induction that:

\underline{{\bf Lemma}}
\begin{equation}
{\cal L}_n \equiv \frac{1}{n !} \left( \frac{i \hbar}{2} \right)^n \hat{{\cal L}}^n A \cdot B.
\end{equation}

\underline{{\bf Proof:}}

From (26) we get $ {\cal L}_0 = AB$ and
$$
{\cal L}_1 = \frac{i \hbar}{2} \sum_{i,j=1}^{2N} \frac{\partial A}{\partial O_i} \frac{\partial B}{\partial O_j} \left\{ O_i, O_j \right\} =\frac{i \hbar}{2} 
\sum_{i=1}^{N} \left(\frac{\partial A}{\partial q_i} \frac{\partial B}{\partial p_i} - \frac{\partial A}{\partial p_i} \frac{\partial B}{\partial q_i} \right) = 
\frac{i \hbar}{2} \left\{ A, B \right\}.
$$
It is easy to check that:
\begin{equation}
\left\{A, B \right\} = \hat{{\cal L}} A \cdot B.
\end{equation}
And so $ {\cal L}_1 =  i \hbar /2  \hat{{\cal L}} A \cdot B$, in agreement with (29), (30). 

Let us now assume that (30) holds for some $n$. We then have from (26):
$$
\begin{array}{c}
{\cal L}_{n+1} = \frac{1}{(n+1) !} \sum_{1 \le i_1, \cdots , i_{n + 1} \le 2N} \sum_{1 \le j_1, \cdots , j_{n + 1} \le 2N} \left( \frac{i \hbar}{2}
\right)^{n+1}  \frac{\partial^{n + 1} A}{\partial O_{i_1} \cdots \partial O_{i_{n + 1}}}  \times \\
\\ 
\times \frac{\partial^{n + 1} B}{\partial O_{j_1} \cdots \partial O_{j_{n + 1}}}  
\times \left\{ O_{i_1} , O_{j_1} \right\} \cdots \left\{ O_{i_{n + 1}} , O_{j_{n + 1}} \right\}  = \\
\\
=  \frac{1}{(n+1) !} \left( \frac{i \hbar}{2} \right)^{n+1} \sum_{1 \le i_1, \cdots , i_n \le 2N} \sum_{1 \le j_1, \cdots , j_n \le 2N}  \left\{ O_{i_1} , O_{j_1} 
\right\} \cdots \left\{ O_{i_n} , O_{j_n} \right\} \times \\
\\
\times \left\{\frac{\partial^n A}{\partial O_{i_1} \cdots \partial O_{i_n}} ,  \frac{\partial^n B}{\partial O_{j_1} \cdots \partial O_{j_n}} \right\} =\\
\\
=  \frac{1}{(n+1) !} \left( \frac{i \hbar}{2} \right)^{n+1} \sum_{1 \le i_1, \cdots , i_n \le 2N} \sum_{1 \le j_1, \cdots , j_n \le 2N}  \left\{ O_{i_1} , O_{j_1} 
\right\} \cdots \left\{ O_{i_n} , O_{j_n} \right\} \times \\
\\
\times \left[ \hat{{\cal L}} \left(\frac{\partial^n A}{\partial O_{i_1} \cdots \partial O_{i_n}} \cdot  \frac{\partial^n B}{\partial O_{j_1} \cdots \partial O_{j_n}} \right) 
\right].
\end{array}
$$
In the last step we used (31). Since $\left\{ O_i, O_j \right\}$ are c-numbers, they commute with the operator $\hat{{\cal L}}$, and we get:
$$
\begin{array}{c}
{\cal L}_{n+1} =  \frac{1}{n+1} \frac{i \hbar}{2} \hat{{\cal L}} \left[ \frac{1}{n !} \left( \frac{i \hbar}{2} \right)^n \sum_{1 \le i_1, \cdots , i_n \le 2N} 
\sum_{1 \le j_1, \cdots , j_n \le 2N} \right.\\
\\
 \left. \left\{ O_{i_1} , O_{j_1} \right\} \cdots \left\{ O_{i_n} , O_{j_n} \right\} \times \frac{\partial^n A}{\partial O_{i_1} \cdots \partial O_{i_n}} \cdot 
\frac{\partial^n B}{\partial O_{j_1} \cdots \partial O_{j_n}} \right] = \\
\\
=   \frac{1}{n+1} \frac{i \hbar}{2} \hat{{\cal L}} \left[{\cal L}_n \right] = \frac{1}{(n+1)!} \left(\frac{i \hbar}{2} \right)^{n+1} \hat{{\cal L}}^{n+1} A \cdot B,
\end{array}
$$
in agreement with (30).  In summary, we proved the following theorem:

\underline{{\bf{Theorem 1}}}: Let $\hat A, \hat B \in \hat{{\cal A}}$ and $A= V_S(\hat A), B=V_S(\hat B)$. The dequantization of the product $\hat A \cdot \hat B$ is given by:
\begin{equation}
V_S( \hat A \cdot \hat B) \equiv A * B = \exp \left( \frac{i \hbar }{2} \hat{{\cal L}} \right) A \cdot B.
\end{equation}
This product is more commonly found in the form:
\begin{equation}
A*B = A \exp \left( \frac{i \hbar}{2} \hat{{\cal J}} \right) B,
\end{equation}
where $\hat{{\cal J}}$ is the Janus operator:
\begin{equation}
\hat{{\cal J}} \equiv \sum_{i=1}^N \left( \frac{ {\buildrel { \leftarrow}\over\partial}}{\partial q_i} \frac{ {\buildrel { 
\rightarrow}\over\partial}}{\partial p_i} -  \frac{{\buildrel { \leftarrow}\over\partial}}{\partial p_i}  \frac{{\buildrel { 
\rightarrow}\over\partial}}{\partial q_i} \right).
\end{equation} 

This product has the following properties for ($A,B,C \in {\cal A};$  $a, b \in {\cal C})$:

\begin{equation}
\begin{array}{l l}
1) & \mbox{Linearity: } (a A + b B ) * C = a (A * C) + b ( B * C),\\
2) & A *_{\hbar} B = B *_{- \hbar} A,\\
3) & \mbox{Associativity: } (A * B) * C= A * ( B * C),\\
4) & \mbox{Identity: } A * 1 = 1 * A = A,\\
5) & 
 p_i * p_j = p_i \cdot p_j; \qquad q_i * q_j = q_i \cdot q_j ; \qquad q_i * p_j = q_i \cdot p_j + \frac{i \hbar}{2} \delta_{ij}\\
6) & (A * B)^* = B^* * A^*
\end{array}
\end{equation}
Property 2) means that changing the order of the variables $A$, $B$ is tantamount to performing the substitution $\hbar \to - \hbar$ in formula (33). This can be proved
immediately  by substituting the identity:
\begin{equation}
B \hat{{\cal J}}^n  A = (-1)^n A \hat{{\cal J}}^n B,
\end{equation}
in equation (30).
Property 3) is a trivial consequence of dequantizing the product of three operators. The remaining properties are straightforward to prove using the 
formula (33).

\subsection{The bracket $\left[ , \right]_M$}

\underline{{\bf{Theorem 2}}}: The classical bracket $\left[A , B \right]_M \equiv V_S( \left[ \hat A, \hat B \right])$ is given by:
\begin{equation}
\left[A , B \right]_M = 2 i A \sin \left( \frac{\hbar}{2} \hat{{\cal J}} \right) B,
\end{equation}
for any $\hat A , \hat B \in \hat{{\cal A}}$ and $A = V_S( \hat A)$, $B=V_S( \hat B)$.

\underline{{\bf{Proof}}}:
$$
\begin{array}{c}
\left[A , B \right]_M \equiv  V_S( \left[ \hat A , \hat B \right]) =  V_S( \hat A \hat B ) -  V_S ( \hat B \hat A)= A * B -B * A = 
A \left( e^{\frac{i \hbar}{2} \hat{{\cal J}}} - e^{- \frac{i \hbar}{2} \hat{{\cal J}}} \right)  B = \\
\\
= \sum_{n=0}^{\infty} \frac{1}{n!} \left( \frac{i \hbar}{2} \right)^n \left[ 1 - (-1)^n \right] A \hat{{\cal J}}^n  B = 2i A \sin \left( \frac{\hbar}{2} 
\hat{{\cal J}} \right) B,
\end{array}
$$
where we used eq.(21) and property 2) in eq.(33). This expression is the celebrated Moyal bracket, \cite{moyal}. This formula might appear awkward at first 
sight. One would expect $V_S( \left[\hat A, \hat B \right]) = i \hbar \left\{A, B
\right\}$. To order $\hbar$, we have:
\begin{equation}
\begin{array}{l l}
V_S(\hat A \cdot \hat B) = & A \left( 1 + \frac{i \hbar}{2} \hat{{\cal J}} + {\cal O} (\hbar^2) \right)  B = A \cdot B +  \frac{i \hbar}{2} \left\{A, B 
\right\}  + {\cal O} (\hbar^2),\\
& \\
V_S( \left[\hat A, \hat B \right])= & 2 i A \left( \frac{\hbar}{2} \hat{{\cal J}} + {\cal O} (\hbar^3) \right) B = i \hbar \left\{A, B \right\}  + 
{\cal O} (\hbar^3).
\end{array}
\end{equation}
To this order we do indeed recover the Poisson bracket. The
Moyal bracket has the following properties for $A, B,C \in {\cal A}$; $a, b \in {\cal C}$:
\begin{equation}
\begin{array}{l l}
1) & \mbox{Linearity: } \left[a A + b B , C \right]_M = a \left[A , C \right]_M + b \left[ B, C \right]_M,\\
2) & \mbox{Antisymmetry: } \left[B , A \right]_M = - \left[A , B \right]_M ,\\
3) & \mbox{Jacobi identity: } \left[ \hspace{0.2cm} \left[A , B \right]_M , C \hspace{0.2cm} \right]_M + \left[ \hspace{0.2cm}
  \left[B , C \right]_M, A \hspace{0.2cm} \right]_M + \left[ \hspace{0.2cm}  \left[C , A \right]_M, B \hspace{0.2cm} \right]_M =0 ,\\
4) & \mbox{Product (Leibnitz) rule: }  \left[A * B, C \right]_M =  A * \left[B , C \right]_M + \left[A , C \right]_M * B,\\
5) & \mbox{Structure constants :} \left[q_i , q_j \right]_M = \left[p_i , p_j \right]_M =0; \qquad \left[q_i , p_j \right]_M = i \hbar \delta_{ij}.
\end{array}
\end{equation}

All these results follow immediately from the properties (35) of the product $*$ and from (21).

\subsection{Dynamical Evolution}

The time evolution of a quantum operator $\hat A (t)$ is given by:
\begin{equation}
\hat A (t) = \sum_{n=0}^{\infty} \frac{1}{n !} \left( \frac{it}{\hbar} \right)^n \left[ \hat H , \left[ \hat H , \left[ \cdots , \left[ \hat H , \hat A \right] \cdots
\right] \right] \right],
\end{equation}
where $\hat H$ is the Hamiltonian. If $V_S (\hat H) = H$, then the previous equation yields upon dequantization:
\begin{equation}
A (t)= \sum_{n=0}^{\infty} \frac{1}{n !} \left( \frac{it}{\hbar} \right)^n \left[\hspace{0.2cm}  H , \left[\hspace{0.2cm}  H , \left[ \cdots , 
\left[\hspace{0.2cm}  H ,  A \hspace{0.2cm} \right]_M \cdots \right]_M \hspace{0.2cm} \right]_M \hspace{0.2cm} \right]_M,
\end{equation}
As a consequence of (41), the observable $A$ obeys the differential equation:
\begin{equation}
\dot A (t) = \frac{i}{\hbar}  \left[H , A (t) \right]_M.
\end{equation}
Notice that this equation could also be obtained by dequantizing the original quantum dynamical equation for the observable $\hat{A}(t)$.
In particular we have:
\begin{equation}
\left\{
\begin{array}{l r}
\dot q_i (t) = \frac{\partial H}{ \partial p_i} & \\
& \qquad i=1, \cdots , N.\\
\dot p_i (t) = - \frac{\partial H}{ \partial q_i} &
\end{array}
\right.
\end{equation}
Notice that these equations look exactly like the ones in traditional classical mechanics (i.e. with Poisson brackets). However, this similarity 
is misleading. Indeed, $H$ does not look exactly like the traditional classical Hamiltonian. Rather, all products of variables are replaced by 
the product $*$.

Let us now study some of the properties of the new theory:\\
1) Time evolution is generated by an unitary transformation. From the quantum theory, we have:
\begin{equation}
\left\{
\begin{array}{l}
\hat A (t) = \hat U (t )^{-1} \hat A(0) \hat U (t),\\
\\
i \hbar \frac{\partial \hat U}{\partial t} = \hat H \hat U, \qquad \hat U(0) = 1.
\end{array}
\right.
\end{equation}
Applying the map $V_S$:
\begin{equation}
\left\{
\begin{array}{l}
A (t) = U (t )^{-1} *  A(0) * U (t),\\
\\
i \hbar \frac{\partial U}{\partial t} = H *  U, \qquad U(0) = 1,
\end{array}
\right.
\end{equation}
where we used the fact that $V_S( \hat U^{-1} \hat U) = V_S( U^{-1}) * V_S( \hat U)=1$, and therefore $V_S(\hat U^{-1}) = V_S( \hat
U)^{-1}$.

Moreover, the classical quantity $U (t)$ is unitary in the sense that $U^{-1} = U^*$ ($\hat U^{-1} = \hat U^{\dagger}  \Rightarrow 
V_S(\hat U^{-1}) = V_S( \hat U^{\dagger}) \Rightarrow V_S(\hat U)^{-1} = V_S ( \hat U)^*$) and thus $U^* * U = 1$. 

Substituting (45) into (42), we can verify explicitly that it provides a solution to the equations of motion.\\
2) All unitary transformations generate canonical transformations. In fact: 
\begin{equation}
\begin{array}{c}
U^{-1} * \left[ A, B \right]_M * U = U^{-1} *  A *  B * U - U^{-1} *  B * A  * U =\\
\\
= U^{-1} *  A * U * U^{-1} *  B * U - U^{-1} *  B * U * U^{-1} * A  * U =\\
\\
= \left[U^{-1} * A * U , U^{-1} * B * U  \right]_M,
\end{array}
\end{equation}
and so the bracket structure is preserved under the action of $U$. In particular, time evolution is a canonical transformation. \\
3) The limit $\hbar \to 0$ of symmetric classical mechanics is standard classical mechanics. Indeed, the identities,
\begin{equation}
\left\{
\begin{array}{l r}
\lim_{\hbar \to 0} A * B = A \cdot B, & \\
& \forall A,B \in {\cal A}\\
\lim_{\hbar \to 0} \frac{1}{i \hbar} \left[ A, B \right]_M = \left\{A, B \right\}, & 
\end{array}
\right.
\end{equation}
can be checked immediately from the expansions (38). Using these limits in eqs.(41) and (42), we recover the standard version of classical
mechanics.\\
4) Symmetric classical mechanics fully validates the use of expansion (4). In fact, if $A(0) = V_S \left( \hat A (0) \right))$, then $A(t)$
(obtained by solving (42)) is given by $V_S \left( \hat A (t) \right)$. Therefore expansion (4) is exactly valid for $\hat A (t) - A(t)$, where
$A(t)$ is the prediction of symmetric classical mechanics for the time evolution of the observable $A(0)$, and thus all the results concerning the
consistency between the classical and quantum predictions are exactly valid if the classical predictions are those of symmetric classical
mechanics.

\subsection{Symmetric Quantization}

Finally, we shall define a quantization prescription for symmetric classical systems:

\underline{{\bf Definition 4:}} {\bf Symmetric Quantization}\\
The symmetric quantization map $\Lambda_S: {\cal A} \to \hat{{\cal A}}$ is the Lie algebra isomorphism defined by the following rules:\\
1) $\Lambda_S$ is linear,\\
2) $\Lambda_S (A * B) = \Lambda_S (A) \cdot \Lambda_S( B).$\\
It satisfies the following properties:\\
1) $\Lambda_S$ maps the identity to the identity:
$$
\Lambda_S (A )= \Lambda_S (A * 1 ) = \Lambda_S (A ) \cdot \Lambda_S (1 ) \Rightarrow \Lambda_S (1 ) = 1.
$$
2) It is the inverse map of $V_S$:
$$
\Lambda_S \left( O_{i_1} O_{i_2} \cdots O_{i_k} \right) =
\Lambda_S \left( (O_{i_1} *  O_{i_2} * \cdots * O_{i_k})_+ \right) =
\left( \Lambda_S ( O_{i_1}) \Lambda_S (O_{i_2}) \cdots  \Lambda_S ( O_{i_k}) \right)_+
$$
where in the last step we used rule 2) from the definition. 
The previous identity together with property 1) proves that $\Lambda_S \circ V_S = 1$.\\
3) A trivial consequence of rule 2) is the following:
$$
\Lambda_S \left( \left[ A,B \right]_M \right) = \left[ \Lambda_S(A),\Lambda_S(B) \right].
$$

\section{Conclusions}

In this paper we presented an alternative formulation of classical physics. The new theory was named symmetric classical mechanics. Its properties
were studied thoroughly and, most important, it was shown that symmetric
classical mechanics is the exact semiclassical limit of quantum mechanics for an arbitrary quantum system with a set of initial data satisfying
the classicality criterion presented in section 2. In other words, the time evolution of a general quantum system is $M$-order consistent with
the predictions of symmetric classical mechanics, provided the initial data for the two formulations are $M$-order classical. Notice that this
property is not completely satisfied by standard classical mechanics.

Clearly, symmetric classical mechanics is not the only possible alternative framework for classical mechanics. The entire set of properties of the new theory and in particular its consistent canonical structure are a direct consequence of the definition of the dequantization map
or, to go even further, of the choice of a basis for the algebra of quantum observables. Therefore, all the results presented in this paper can
be reformulated for other dequantization maps, providing in this fashion other, possibly more interesting, descriptions of classical physics,
\cite{ou}.

The motivation to develop the theory of symmetric classical mechanics was threefold: 1) The first and most important motivation is theoretical. Symmetric 
classical mechanics provides a new perspective over the problem of the semiclassical limit of quantum mechanics. Firstly, because it proves that the semiclassical limit might be correctly described (and even more accurately) by another fully consistent dynamical structure and not just by 
classical mechanics. Secondly, because it clarifies the role of the limit $\hbar \to 0$ in deriving the semiclassical limit of quantum mechanics. Symmetric 
classical mechanics provides a description of the semiclassical limit in which no assumption is made about the magnitude of the Planck constant. Its validity 
rests exclusively upon a number of conditions that should be satisfied by the initial data wave function. In other words, symmetric classical mechanics would still 
provide a valid description of dynamics in a world with a huge Planck constant. As a side result we see that standard classical mechanics can be seen as a 
second limit of quantum mechanics when the set of initial data of the dynamical system satisfies some classicality conditions and the Planck constant can be regarded 
as being of neglectable magnitude. This result corroborates the argument and the results of \cite{nuno1}. 
2) On the other hand, symmetric classical mechanics is formulated in terms of the Moyal bracket and this bracket also provides the dynamics of the 
Moyal-Weyl-Wigner formulation of quantum mechanics. Therefore, it comes as no surprise that
when compared to standard classical mechanics, symmetric classical mechanics displays a clearer relation with quantum mechanics. The quantization map
from symmetric classical mechanics to quantum mechanics is one-to-one and there are thus no order problems in the quantization of a symmetric
classical system. This property provides a new approach for the analysis of the order ambiguities in quantum mechanics. This analysis can
now be enforced at the level of the original classical theory. 3) The last motivation concerns the problem of developing a consistent theory of
coupled classical-quantum dynamics \cite{maddox,boucher,nuno2,anderson,diosi}. We expect that the symmetric dequantization map might be consistently extended to the case where
the purpose is to dequantize only one sector of the original quantum theory. If this is the case and if the extended dequantization map
preserves the original set of properties, then it is trivial to obtain a consistent formulation (i.e. a true Lie bracket structure) of hybrid classical-quantum dynamics.

\subsection*{Acknowledgments}

We would like to thank Jo\~{a}o Marto for 
several suggestions made through the present work.
This work was partially supported by the grants 
ESO/PRO/1258/98 and CERN/P/Fis/15190/1999.

\end{document}